\documentclass[pra,showpacs,twocolumn,superscriptaddress,floatfix,aps]{revtex4-2}

\usepackage[dvips]{graphics}
\usepackage{color}
\usepackage{float}
\usepackage{natbib}
\usepackage[dvipsnames]{xcolor}
\setcitestyle{numbers}
\setcitestyle{square}
\usepackage{graphicx}
\usepackage{bm}
\usepackage{amssymb}
\usepackage{amsmath}
\usepackage{mathtools}
\usepackage{dcolumn}  
\usepackage[colorlinks, linkcolor=Blue,citecolor=Blue,urlcolor=Blue]{hyperref}

\usepackage[all]{hypcap} 
\usepackage{physics}
\usepackage{enumerate}
\usepackage{braket}       
\usepackage{overpic}
\usepackage{amsfonts}
\usepackage{dsfont}
\usepackage[mathscr]{eucal}
\usepackage{hyperref}
\usepackage{setspace}

\begin{document}
\title{Fractional Shapiro steps in a Cavity-Coupled Josephson ring condensate}
\author{Nalinikanta Pradhan}
\affiliation{Department of Physics, Indian Institute of Technology, Guwahati 781039, Assam, India}

\author{Rina Kanamoto}
\affiliation{Department of Physics, Meiji University, Kawasaki, Kanagawa 214-8571, Japan}

\author{M. Bhattacharya}
\affiliation{School of Physics and Astronomy, Rochester Institute of Technology, 84 Lomb Memorial Drive, Rochester, New York 14623, USA}
\author{Pankaj Kumar Mishra}
\affiliation{Department of Physics, Indian Institute of Technology, Guwahati 781039, Assam, India}

\date{\today}

\begin{abstract} 
The Josephson effect presents a fundamental example of macroscopic quantum coherence as well as a crucial enabler for metrology (e.g. voltage standard), sensing (e.g. Superconducting Quantum Interference Device) and quantum information processing (Josephson qubits). Recently, there has been a major renewal of interest in the effect, following its observation in Bose, Fermi, and dipolar atomic condensates, in exciton-polariton condensates, and in momentum space. We present theoretically a nondestructive, \textit{in situ} and real time protocol for observing the AC and DC Josephson effects including integer (recently observed in cold atoms) and fractional  (hitherto unobserved in cold atoms) Shapiro steps, using a ring condensate coupled to an optical cavity. Our analysis presents a metrology standard that does not require measurmement of atomic number and that challenges the conventional wisdom that quantum computations cannot be observed without being destroyed. Our results have implications for the fields of atomtronics, sensing, metrology and quantum information processing.     
\end{abstract}

\flushbottom

\maketitle


\textit{Introduction:} 
The Josephson effect~\cite{JOSEPHSON1962251}, first proposed to describe the emergence of particle current between weakly coupled superconductors~\cite{AndersonPRL1963}, has been a topic of immense interest in the condensed matter and quantum optics communities both as an exemplar of macroscopic quantum coherence as well as an enabler of applications in metrology and quantum information processing~\cite{Petley1969, BADIANE2013840,jeanneret2009application,bobkova2022magnetoelectric,kockum2019quantum,RevModPhys.73.357,kim2025josephsonjunctionsagequantum}. Subsequently, the effect was observed in an array of reservoirs connected via submicron apertures in superfluid $^3$He~\cite{he3quantum1997}. There has been a renewed interest in the topic from the ultracold atoms community following the realization of the bosonic Josephson junction ~\cite{PRL2005,gati2006,levy2007acdc,fantoni2000,experiment1DJJ}, which differs from its superconductor and superfluid counterparts due to its high experimental tunability and ease of theoretical characterization. Recently, Josephson effects have also been observed in supersolids~\cite{BiagioniNature2024,donelliselfinducedJosephsoneffectindipolarbec} and in momentum space~\cite{mukhopadhyayPRL2024momentumspaceJosephsonEffect,dutta2025rotationmediatedbosonicjosephsonjunctions} and studied theoretically in various configurations using the Gross-Pitaevskii equation and Bogoliubov analysis~\cite{smerzi1997,Giovanazzi2000,Meier2001,Burchianti2017,gupta2025theoreticaldescription}. A key finding in this context is the emergence of Shapiro steps—a hallmark of phase synchronization between two reservoirs and an externally driven Josephson junction~\cite{shapiro1963}. These have been observed in both Bose and Fermi systems~\cite{singh2024,bernhart2024observation,delpace2024shapirofermi,singh2025parametricamplifier}, representing the accession of the voltage standard to the platform of ultracold quantum gases.  But the observation of fractional Shapiro steps, already performed in superconducting systems~\cite{Vanneste1988,EARLY1995,CHERN1997,Sellier2004,Ueda2020,Yao2021,zhang2025}, has not yet been made in ultracold quantum gas systems.


Specifically, inspired by the Superconducting Quantum Interference Devices (SQUIDs)~\cite{SQUIDbook}, Atomic Quantum Interference Devices (AQUIDs)~\cite{experiment1DJJ,eckel2014hysteresis,chien2015quantum,ryu2020quantum} have been developed using a Bose-Einstein condensate (BEC) confined to an annular trap, which naturally supports the atomic supercurrent while preserving the coherence of the condensate phase~\cite{ringtrap,RyuPRL2007,MoulderPRA2012,TAAP2012,pandey2019hypersonic,wright2013driving,MurrayPRA2013,CormanPRL2014}. An AQUID typically consists of one or more Josephson junctions, in the form of optical tunneling barriers, allowing observation of phase slips~\cite{wright2013driving, Kiehn2022}, the DC-AC Josephson effect~\cite{levy2007acdc,experiment1DJJ}, quantum interference of currents~\cite{readout2018,ryu2020quantum,Kiehn2022}, parity-protected qubits~\cite{qubit2024}, and stabilization of persistent currents~\cite{pezze2024stabilizing}. Recently, the realization of the Josephson ring junction in an exciton-polariton condensate has opened the door to applications at room temperature~\cite{voronova2025exciton}. 

These demanding applications of cold-atom Josephson systems in general and AQUIDs in particular require precise information on the motional dynamics of the condensate. To date, 
all realized (see above) and proposed \cite{readout2018} detection of current states in cold-atomic Josephson junctions has been totally destructive, requires time-of-flight expansion of the atomic gas, and cannot be carried out in real time. These restrictions require repeated sample preparation, expose the system to variations in initial conditions, slow down metrological measurements, and forbid checks during quantum computations.

\begin{figure}[t]
\begin{center}
    \includegraphics[width=1\linewidth]{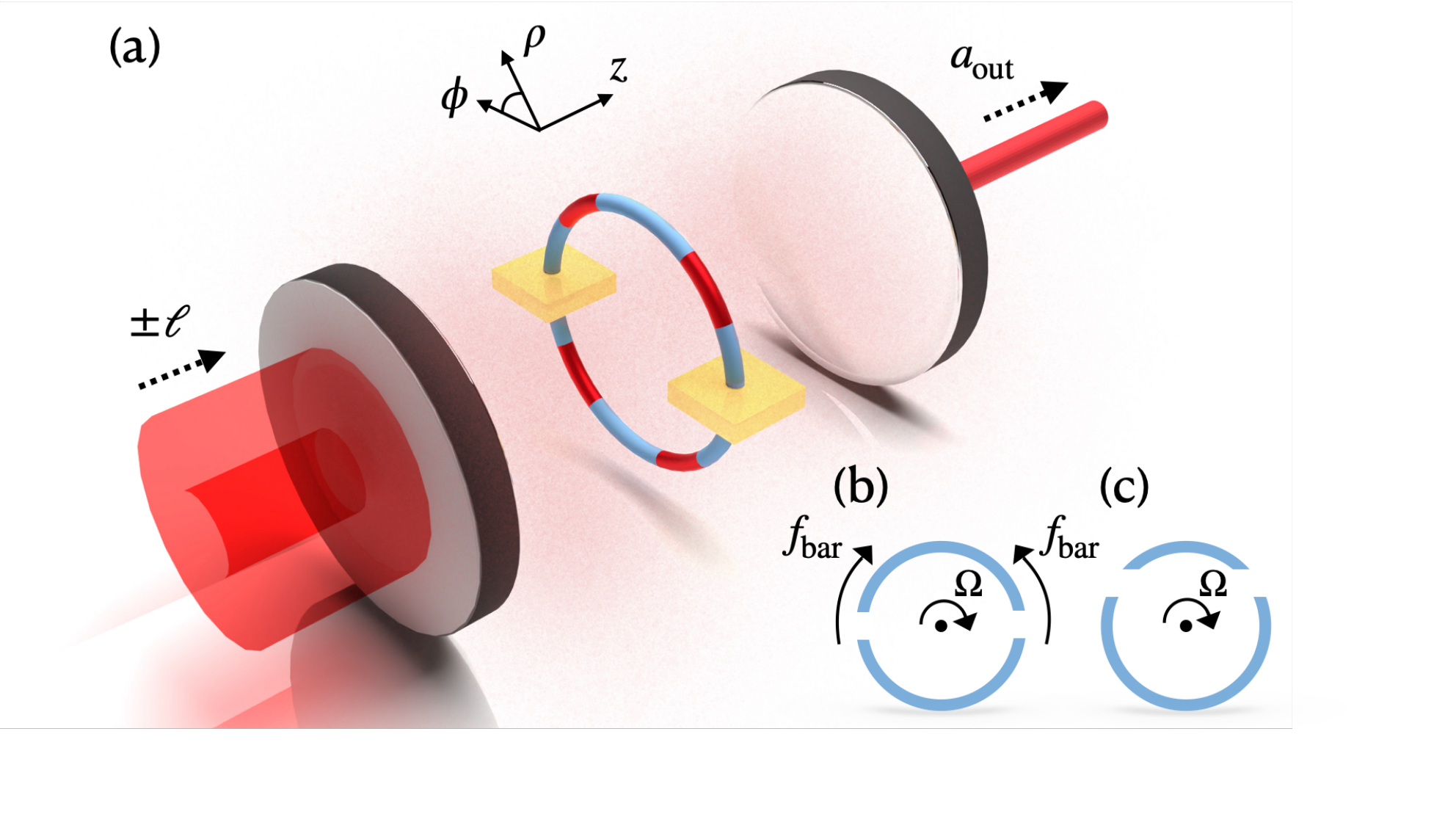} \\

\end{center}
\vspace{-8pt}
\caption{(a) Schematic diagram of proposed experimental setup. A ring-trapped BEC with two Josephson junctions (JJs) is placed inside an optical cavity.  The Josephson junction dynamics is probed by two Laguerre-Gaussian (LG) beams carrying orbital angular momenta $\pm \ell \hbar $ that form an annular lattice (red shaded region). $a_{out}$ denotes the optical field transmitted by the cavity. (b) Initial setup with the JJs that are placed symmetrically across the ring, resulting in two half rings. The JJs are rotated at angular frequency $\Omega$, and simultaneously moved towards each other with velocity $f_{\mathrm{bar}}$ up to a time interval $t_{\mathrm{bar}}$ to generate a chemical potential difference between the two regions. (c) Final positions of the JJs beyond $t_{\mathrm{bar}}$. }   

\label{fig:setup}
\end{figure}

To remove all these restraints, in this Letter, we theoretically propose a nondestructive, \textit{in situ} and real-time method of measuring Josephson oscillations. Our protocol is based on our previous work on the detection of persistent currents and solitons~\cite{KumarPRL2021,pradhan2024cavity,pradhan2024ring}, later extended to the detection of superfluid drag~\cite{pradhan2025AndreevBashkin}, and exploits the sensitive techniques of cavity optomechanics~\cite{AspelmeyerRMP2014,BrenneckeScience2008}. In this work We consider the much more complicated case of a ring condensate with Josephson junctions coupled to an optical cavity driven by orbital angular momentum-carrying beams. Inspite of the highly multimode nature of the dynamics, we show how our technique can yield clean signatures, in the cavity transmission, of the DC and AC Josephson effects as well as the Shapiro steps. Although for reasons of simplicity we restrict our present demonstration to AQUIDs, our results can readily be generalized to other condensate and cavity geometries.   

\textit{Physical configuration:} The method requires an annularly magnetically trapped condensate with optically enforced Josephson junctions to be kept at the center of a high-finesse optical cavity, as shown in Fig.~\ref{fig:setup}. Driving the cavity with a superposition of two Laguerre-Gaussian beams of orbital angular momentum (OAM) $\pm \ell\hbar$~\cite{leach2010}, leads to a weak angular optical lattice~\cite{NaidooAPB2012}. When a rotating matter wave with winding number $L_p$ interacts with this lattice, it undergoes Bragg diffraction into rotational states $L_p \pm 2 n \ell$, where $n$ is an integer. These currents serve as the mechanical modes optomechanically interacting with the intra-cavity field~\cite{BrenneckeScience2008,KumarPRL2021}. The cavity field is in turn modulated at the condensate's side-mode as well as Josephson oscillation frequencies, which can be obtained by a heterodyne detection of the cavity-output field~\cite{AspelmeyerRMP2014}.







 
\textit{Theoretical Model:} We numerically model the proposed experimental setup by considering a one-dimensional ring BEC of $^{23}$Na atoms with multiple Josephson junctions placed inside an optical cavity. The coupled dynamics of the condensate wave function $\psi (\phi,t)$ and cavity field amplitude $\alpha$ are described by the following set of non-dimensionalised equations \cite{AspelmeyerRMP2014, BrenneckeScience2008, KumarPRL2021, pradhan2024cavity, pradhan2024ring}
\begin{equation}
\begin{aligned}
    (i-\Gamma)&\frac{d\psi}{d\tau} = \biggl[( -i \frac{d}{d\phi} - \Omega')^2 + \frac{U_0}{\omega_\beta}  |\alpha(\tau)|^2 \cos^2\left({\ell\phi}\right) \\ & + V_{\mathrm{bar}} (\phi,\tau)  -\mu  + \mathcal{G} | \psi (\phi,\tau)|^2 \biggr] \psi + \chi(\phi,\tau),
    \label{Eq:bec}    
\end{aligned}
\end{equation}
\begin{equation}
\begin{aligned}
    i\frac{d\alpha}{d\tau} = \biggl\{ - \biggl[\Delta_c &- U_0 \langle \cos^{2}\left(\ell\phi\right)\rangle_{\tau} + i \frac{\gamma_{0}}{2}\biggr] \alpha  + i\eta \biggr\} \omega_\beta^{-1} \\ &+ i \sqrt{\gamma_0} \omega_\beta^{-1} \alpha_{in}(\tau). 
    \label{Eq:cavity}
\end{aligned}
\end{equation}
We employ the dimensionless formulation by scaling energy as $\hbar \omega_{\beta} = \hbar^2 / 2 m R ^2$ and time as $\tau = \omega_\beta t\;$. Here $m$ is the mass of the atom and $R$ the radius of the ring trap, also used as the unit of length ($\phi = x / R$). The condensate wave function is normalized to the total number of atoms via $\int |\psi(\phi,t)|^2 \, d\phi = N$. The terms of Eqs.~(\ref{Eq:bec}) and (\ref{Eq:cavity}) are described in detail below. The first term on the right-hand side of Eq.~(\ref{Eq:bec}) represents the rotational kinetic energy of the atoms in the frame co-rotating with barriers, with rotation frequency $\Omega = \Omega' \times (2\omega_\beta)$. The second term describes the optical lattice potential that couples the condensate to the cavity field with strength $U_0 = g_a^2/ \Delta_a$, where $g_a$ is the single-atom-photon coupling strength and $\Delta_a$ is the detuning of the optical field from the atomic transitions. The third term represents the optical barrier potential, which can be created by ``painting'' a repulsive focused beam onto the condensate. It is modeled as \cite{pezze2024stabilizing}
\begin{eqnarray}
	V_{\text{bar}}(\phi,\tau) = V_0 \left[ e^{-2 \left( \frac{\phi - \phi_{b1}}{\sigma} \right)^2} + e^{-2 \left( \frac{\phi - \phi_{b2}}{\sigma} \right)^2} \right],
\end{eqnarray}
where $\phi_{b1}$ and $\phi_{b2}$ are the positions of two barriers that act as weak links, their heights and widths of the barriers given by $V_0 = 0.2\, \mu_0/\hbar \omega_\beta$ and, $\sigma = 1.2 \, \xi$ respectively, where $\mu_0$ is the chemical potential of the condensate without barriers and $\xi$ is the healing length of the condensate. A similar experimental setup has previously been demonstrated to observe the Josephson effect in a ring BEC \cite{experiment1DJJ,pezze2024stabilizing}. The application of the two optical barriers leads to the formation of two half-ring condensates, and the total number of atoms is distributed between the two rings as $N_j = \int\int_{S_j} |\psi|^2 \,d \phi,$ where the integration is performed over the respective regions. The first region corresponds to $ S_1: ( \phi_{b1} \leq \phi < \phi_{b2} )$, and the second region to $ S_2 : ( \phi < \phi_{b1}  \, \mathrm{or}\, \phi > \phi_{b2}) $. The fourth term on the right hand side of Eq.~(\ref{Eq:bec}) denotes the chemical potential of the condensate, which has been updated in each time step to conserve the norm of the wave function~\cite{mithun2018signatures} in the presence of the dissipation $\Gamma$ and thermal fluctuation $\chi$.  The fifth term accounts for the two-body atomic interaction with strength $\mathcal{G}= g/\hbar \omega_\beta$, where $g=2\hbar\omega_{\rho}a_{s}/R$, with the $s$-wave scattering length $a_s$ and the harmonic trap frequency along the radial direction $\omega_{\rho}$. 

The first term in the right hand side of Eq.~(\ref{Eq:cavity}) represents the detuning of the driving laser field from the cavity resonance frequency ($\Delta_c = \omega_L - \omega_c$ ), which governs the intra-cavity field build-up and thus controls the dynamical response of the system. The second term contains the expectation value of the optical lattice potential, which describes the coupling between the cavity field and the condensate. In the third term, $\gamma_0$ denotes the cavity linewidth and $\eta=\sqrt{P_{in}\gamma_o/\hbar\omega_c}$ is the pump rate, where $P_{in}$ is the input optical power. The final terms of both Eqs.~(\ref{Eq:bec}) and (\ref{Eq:cavity}) account for the thermal fluctuations in the condensate and the optical shot noise in the cavity field, respectively. These stochastic contributions are described by delta-correlated white noise, i.e.~\cite{das2012winding,KumarPRL2021}
\begin{align} 
\langle\xi(\phi,\tau) \, \xi ^ *(\phi',\tau')\rangle  &= \frac{2\,\Gamma \,k_B \,T}{\hbar \omega_\beta} \, \delta(\phi - \phi') \, \delta (\tau - \tau'), \\ 
\langle \alpha_{in}(\tau) \, \alpha_{in}^ *(\tau') \rangle  &= \omega_\beta \, \delta (\tau - \tau'),
\end{align}
where $k_B$ is the Boltzmann constant and $T$ is the temperature of the BEC. 

We have generated the ground state of the ring BEC in the presence of the barrier potential by using the imaginary time evolution method. This results in two half-rings separated by two symmetrically placed Josephson junctions (JJs). To see the Josephson dynamics, we move the JJs towards each other up to a time interval $t_{\mathrm{bar}}$ with velocity $f_{\mathrm{bar}}$. In the rotating frame, we model the barrier motion as $\phi_{b_{1,2}} = \pm(2 \pi f_{\mathrm{bar}} ) t$. To measure the Josephson oscillation frequency, we turn the laser drive on at $t =  t_{\mathrm{bar}} $, and solve the coupled BEC-cavity equation (Eqs.~(\ref{Eq:bec}) and (\ref{Eq:cavity})).


\begin{figure}[!htp]
\begin{center}
\includegraphics[width= 1\linewidth]{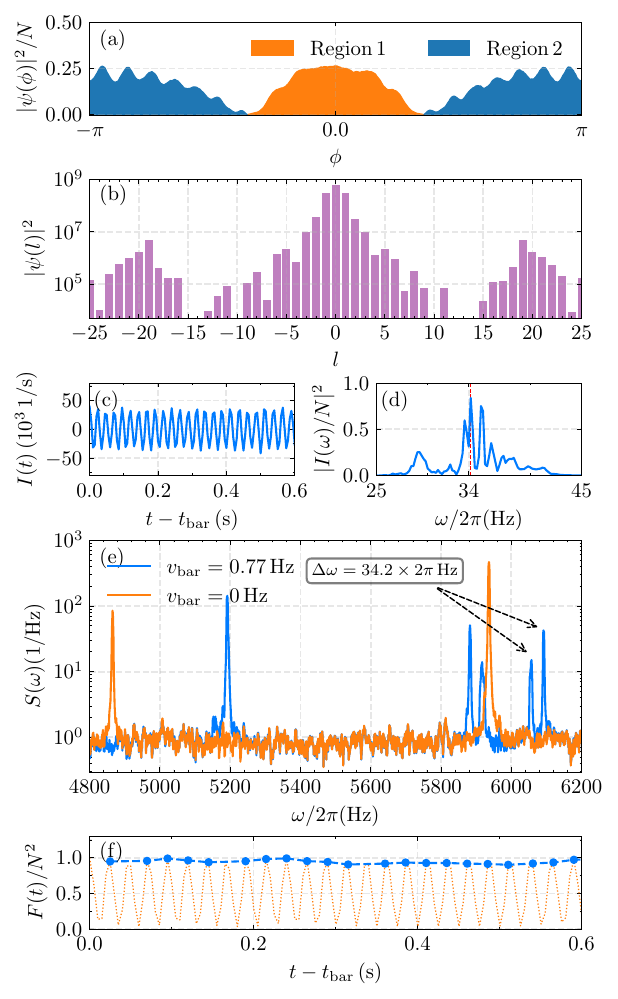} 
\end{center}
\vspace{-8pt}
\caption{ Numerical results obtained after solving Eqs.~\ref{Eq:bec} and \ref{Eq:cavity}: (a-b) Density profile and OAM distributions of the condensate wave function, respectively. (c) Josephson tunnelling current across the barrier. (d) Power spectrum of Josephson tunnelling current. The dominant peak corresponds to the Josephson
oscillation frequency $\omega_{\mathrm{J}}\sim 34.2 \, \mathrm{Hz}$ as shown with red dashed line. (e) Power spectrum of the phase quadrature of the cavity output field. (f) Temporal evolution of fidelity of condensate wave function. The blue dots connect the highest points in the fidelity curve (orange line) that oscillates at the Josephson tunneling frequency $\omega_{\mathrm{J}}$. The parameter values used are $m = 23$ amu, $N =  8000$, $R = 4$ $\mu$m, $f_{\mathrm{bar}} = 0.77 \, \mathrm{Hz}$, $V_0 = 0.2\,\mu_0$, $\sigma = 1.2 \, \xi$, $\xi = 1/\sqrt{\mu_0/\hbar \omega_\beta}$, $t_{\mathrm{bar}}=0.092$ seconds, $\Gamma = 0.0001$, $T = 10$ nK, $P_{\mathrm{in}} = 1 \, \mathrm{pW}$, $ U_0 = 2\pi \times 212$ Hz, $\Tilde{\Delta} = \Delta_0 - U_0 N/2 = -2 \pi \times 173$ Hz, $\Delta_a = 2\pi \times 4.7$ GHz, $\omega_c = 2\pi \times10^{15} $ Hz, $\omega_\rho = \omega_z = 2\pi \times 42$ Hz, $\gamma_{0} = 2\pi\times 2$ MHz and $\Omega' = 0$~\cite{WrightPRL2013, EckelNature2014}.}
\label{fig:fig2}
\end{figure}

\begin{figure*}[!htb]
\begin{center}
\includegraphics[width= 1\linewidth]{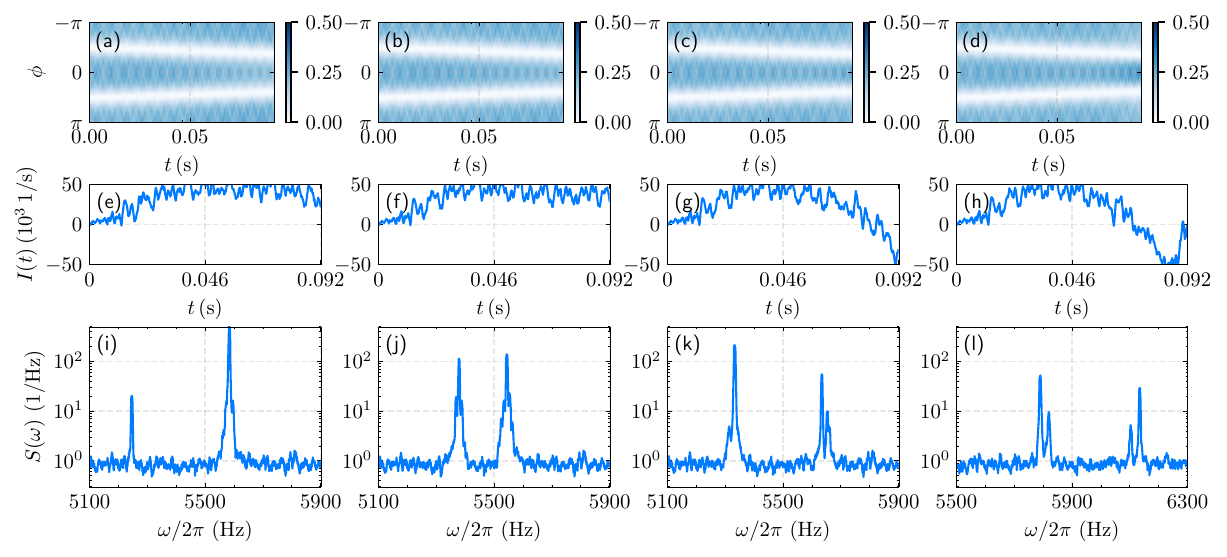} \\

\end{center}
\vspace{-8pt}
\caption{DC-AC Josephson transition [(a)-(d)] Time evolution of condensate density during the barrier movement. [(e)-(h)] Time evolution of particle curent during the junction movement. [(i)-(l)] Power spectra of phase quadrature of the cavity output field vs. the system's response frequency. Here junction velocities are $f_{\mathrm{bar}} = 0.65 \, \mathrm{Hz},\, 0.7 \, \mathrm{Hz}, \, 0.72 \, \mathrm{Hz}$, and $ 0.75 \, \mathrm{Hz}$ respectively. The other set of parameters is the same as used in Fig.~\ref{fig:fig2}.}
\label{fig:fig3}
\end{figure*}


\textit{Detecting Josephson Oscillation frequency:} To demonstrate the detection method, we show the numerical results obtained by solving Eqs.~(\ref{Eq:bec}) and (\ref{Eq:cavity}) for a particular barrier velocity $f_{\mathrm{bar}} = 0.77 \, \mathrm{Hz}$. in Fig.~\ref{fig:fig2}. This barrier movement results in compression of atoms in one region as shown in Fig.~\ref{fig:fig2} (a), and this compression results in a chemical potential difference $\Delta \mu$ between the two regions. Now, the phase difference across the junction ($\Phi = \Phi_2 - \Phi_1$) evolves as $\frac{d\Phi}{dt} = -\frac{\Delta\mu}{\hbar}$ \cite{levy2007acdc,experiment1DJJ}. This triggers the Josephson oscillation in the population imbalance $(n =  N_2 - N_1)$ at the frequency $\omega_J = \Delta \mu / \hbar$. To characterize the condensate's rotational state, we have shown the orbital angular momentum (OAM) distribution in Fig.~\ref{fig:fig2} (b). The dominance of $l = 0,\pm 1$ modes indicates the quantum superposition of persistent current states induced by the moving barriers~\cite{readout2018, Cataldo2024}. Denoting the winding number of these states as $L_p$, they Bragg scatter to $L_p \pm 2\ell$ states in the presence of the weak optical lattice potential \cite{KumarPRL2021}. 
The Josephson tunnelling current flowing across the JJs is calculated by numerically differentiating the population imbalance: $I = \frac{dn}{dt}$ ~\cite{levy2007acdc,experiment1DJJ}. Performing a Fourier transformation of this tunnelling current allows us to calculate the oscillation frequency~\cite{levy2007acdc}, which is $\omega_\mathrm{J}= 34 \times 2\pi$ Hz for this case (See Fig.~\ref{fig:fig2} (d)). Remarkably, as we propose, this Josephson oscillation frequency can be directly obtained from the power spectrum of the phase quadrature of the cavity output field ($S(\omega)=\left|\mathrm{Im}\left[\alpha_{out}(\omega)\right]\right|^{2}$), where $\alpha_{out} (t)$ is calculated numerically by using the input-output relation $\alpha_{out} = -\alpha_{in} + \sqrt{\gamma_0}\alpha$ \cite{AspelmeyerRMP2014}. We emphasize that our method does not require any imaging technique to measure the number of atoms in each region, or destruction of the condensate, as needed in previously available methods for studying Josephson dynamics in ultracold quantum gases~\cite{levy2007acdc,RyuPRL2013,Jendrzejewski2014,ryu2020quantum,bernhart2024observation}. 

We present the power spectrum of the phase quadrature of the cavity output field as a function of the system response frequency for the above-mentioned case in Fig.~\ref{fig:fig2} (e). We see the condensate sideband peaks split for junction velocity $f_{\mathrm{bar}} = 0.77 \, \mathrm{Hz}$, which contrasts with the case when the junctions do not approach each other ($f_{\mathrm{bar}} = 0 \,\mathrm{Hz}$). The reason for this splitting is due to the generation of a chemical potential difference that triggers the Josephson oscillation in the population imbalance. This population oscillation modulates the light field, creating additional sidebands and splitting the peaks in the spectrum by a magnitude $\Delta\omega$ equal to the Josephson frequency $\omega_J$. Importantly, both the spectral peaks show the same magnitude of splitting, highlighting that the splitting originates from a single coherent Josephson tunneling. So,  by observing these splits in the cavity output spectrum, the Josephson oscillation frequency can be detected without destroying the condensate or measuring the atom number. To ascertain the efficacy of our proposed scheme we compute the fidelity of condensate wave function over time, defined as, $F(t) = \int_{-\pi}^{\pi} \left[\psi^{*}(\phi,t)\psi(\phi,0)\right]^2 d\phi$  and present it in Fig.~\ref{fig:fig2} (f). We find that the fidelity remains close to unity, evidencing the non-destructive nature of the measurement technique proposed in this work.

\textit{Detecting DC to AC Josephson transition:} We demonstrate the detection of transition from the DC to AC Josephson regime by varying the junction velocity ($f_{\mathrm{bar}}$), as shown in Fig.~\ref{fig:fig3}. The time evolution of condensate density and particle current during the barrier movement is presented in Figs.~\ref{fig:fig3} (a)-(d) and Figs.~\ref{fig:fig3} (e)-(h), respectively. For barrier velocity $f_{\mathrm{bar}}$ below a critical value $f_{\mathrm{c}}$ the condensate densities in the two regions remain the same due to the coherent tunneling of atoms across the JJs maintaining zero chemical potential difference (Figs.~\ref{fig:fig3} (e),(f)). Beyond $f_{\mathrm{c}}= 0.72\,\mathrm{Hz}$, when the critical current of the JJs is reached, we see the compression (expansion) of atoms in region 1 (region 2) corresponding to $\Delta \mu \neq 0$ (Fig.~\ref{fig:fig3} (d)). This manifests as the Josephson oscillation of population imbalance as seen in Fig.~\ref{fig:fig3} (h).

\begin{figure*}[!htb]
\begin{center}
\includegraphics[width= 1\linewidth]{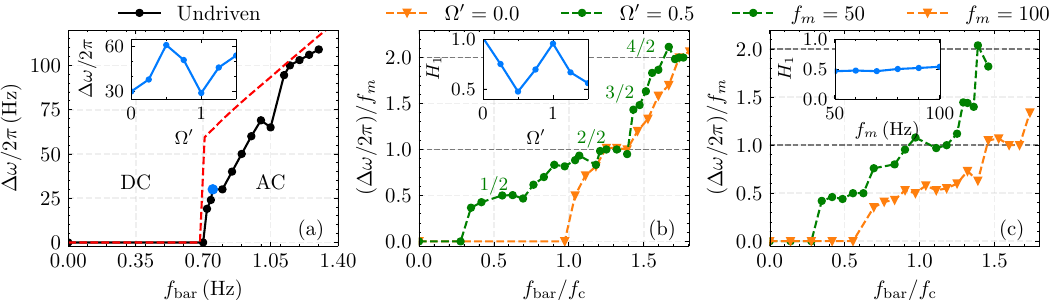} \\

\end{center}
\vspace{-8pt}
\caption{Variation of the amount of peak splitting ($\Delta \omega$) in the cavity output spectra as a function of the barrier velocity ($f_{\mathrm{bar}}$) for (a) $f_{\mathrm{m}} =0\,\mathrm{Hz},\,  \phi_I = 0,\, \Omega' = 0$, (b) $f_{\mathrm{m}} = 60 \,\mathrm{Hz},\, \phi_I = 0.004$, and (c) $\Omega' = 0.5,\,\phi_I = 0.004$. The red dashed line in (a) represents the results obtained from the resistively-capacitively shunted junction (RCSJ) model (See details in supplementary material \cite{supplement}. The inset in (a) shows the oscillation in $\Delta \omega$ with the rotation rate of AQUID for $f_{\mathrm{bar}} = 0.75\, \mathrm{Hz}$. The inset in (b) shows the variation of the first step height $H_1$ with AQUID rotation rate $\Omega'$. The inset in (c) shows the first step height $H_1$ for different modulation frequencies $f_{\mathrm{m}}$. The variation of $\Delta \omega$  as a function of $f_{\mathrm{bar}}$ for these cases are shown in the supplementary material \cite{supplement}. The other set of parameters is the same as used in Fig.~\ref{fig:fig2}. }
\label{fig:fig4}
\end{figure*}

As demonstrated in Fig.~\ref{fig:fig2}, the onset of non-zero chemical potential difference gives rise to a measurable splitting in the peaks in the cavity output spectrum, with the amount of splitting revealing the Josephson oscillation frequency. These spectral features are well above the noise floor, allowing us to measure them precisely. The magnitude of the splitting further increases with an increase in the junction velocity $f_{\mathrm{bar}}$ as seen in Figs.~\ref{fig:fig3} (k),(l), and Fig.~\ref{fig:fig4} (a). Summarizing these observations, we distinguish between the DC and AC Josephson regimes by the presence or absence of peak splitting ($\Delta \omega$) in the cavity output spectrum, respectively. In the inset of Fig.~\ref{fig:fig4}(a), we show the variation of $\Delta \omega$ with the rotation rate $\Omega'$ of AQUID, which oscillates with a period corresponding to one flux quantum $\hbar/mR^2$ (The relevant cavity output spectra are shown in the supplementary material \cite{supplement}). This particular feature demonstrates that our system enables non destructive detection of quantum interference of clockwise and anti-clockwise currents~\cite{readout2018,ryu2020quantum}, which can be effectively utilized as a quantum rotation sensor analogous to a SQUID magnetometer~\cite{SQUIDbook} as well as a qubit in quantum information processing~\cite{AghmalyanNJP2015}.  

\textit{Observing Shapiro steps:} In the bosonic JJ, Shapiro steps appear as plateaus in chemical potential difference ($\Delta \mu$) at integer multiples of external driving frequency~\cite{bernhart2024observation, singh2024}. The external drive is applied through a combination of linear translation (DC current) and periodic modulation (AC current) of junction position, effectively mimicking the external microwave radiation applied to superconducting JJs~\cite{shapiro1963}. To observe the Shapiro steps through our setup (AQUID geometry), we periodically drive JJs' positions along with the linear translations up to $t_{\mathrm{bar}}$. In the frame rotating at $\Omega,$ we model the JJs' positions as $\phi_{b_{1,2}} = \pm (2 \pi f_{\mathrm{bar}} ) t + \phi_i \, \sin(2 \pi f_{\mathrm{m}} t)$,
where $f_{\mathrm{bar}}$ is the junction velocity corresponding to linear translation, $f_{\mathrm{m}}$ is the modulation frequency, and $\phi_i$ is the modulation amplitude. Then we solve Eqs.~(\ref{Eq:bec}) and (\ref{Eq:cavity}) to obtain the cavity output spectra and subsequently, the peak splitting $\Delta \omega$ that acts as the DC voltage in the SQUID. We plot the variation of $\Delta \omega$ for various barrier velocities $f_{\mathrm{bar}}$ in Fig.~\ref{fig:fig4} (b) and see the emergence of Shapiro steps at integer multiples of modulation frequency $f_{\mathrm{m}}$ for $\Omega' = 0$. The width of the Shapiro steps can be further increased by increasing the modulation amplitude $\phi_I$~\cite{singh2024,bernhart2024observation}. However, by rotating the AQUID device (i.e., rotating the barriers), a distinct feature arises, namely fractional Shapiro steps between the integral steps. At half-integer flux quantum, when the system is in a degenerate state of adjacent momentum states~\cite{readout2018}, we can see the plateau at exactly half-integer multiples of the modulation frequency.  This observation is analogous to the occurrence of half-integer Shapiro steps in DC SQUID when it is exposed to an external magnetic field~\cite{Vanneste1988,EARLY1995}.

\textit{Conclusion:} We have proposed and numerically simulated a cavity optomechanical technique for the non-destructive measurement of the Josephson oscillation frequency in an atomtronic quantum interference device. The cavity output spectra reveal the Josephson oscillation frequency precisely as a resolvable splitting in the sidemode peaks, without requiring the challenging measurement of atom number. The onset of the DC to AC Josephson transition and emergence of the integer (recently detected in cold atoms) and fractional (yet to be detected in cold atoms) Shapiro steps are directly observed through these spectral features, offering a clear and experimentally accessible signature of the Josephson tunneling dynamics. In the present work, we have mainly focused on detecting the half-integer Shapiro steps. However, other fractions can be readily accessed with asymmetric barriers or modulation with higher
harmonics. Our approach highlights the potential of cavity optomechanics as a sensitive, non-destructive probe for atomtronic quantum interference devices and facilitates progress in quantum computing and information processing with ultracold atoms. 

\textit{Acknowledgements:} We gratefully acknowledge the supercomputing facilities Param Ishan and Param Kamrupa where all the simulation runs were performed. M.B. thanks the Air Force Office of Scientific Research
(FA9550-23-1-0259) for support. 
R.K. is supported by JSPS KAKENHI (Grant No.25K07190), and JST ERATO (Grant No. JPMJER2302).

\bibliography{citation.bib} 



\clearpage

\widetext

\begin{center}
\textbf{\large Supplementary material: Fractional Shapiro steps in a Cavity-coupled Josephson ring condensate}
\end{center}

\setcounter{equation}{0} \setcounter{figure}{0} \setcounter{table}{0} %
\setcounter{page}{1} \setcounter{section}{0} \makeatletter
\renewcommand{\theequation}{S\arabic{equation}} \renewcommand{\thefigure}{S%
\arabic{figure}} \renewcommand{\bibnumfmt}[1]{[S#1]} \renewcommand{%
\citenumfont}[1]{S#1} \renewcommand{\thesection}{S\arabic{section}}%
\setcounter{secnumdepth}{3}

\renewcommand{\thefigure}{SM\arabic{figure}} \renewcommand{\thesection}{SM
\arabic{section}} \renewcommand{\theequation}{SM\arabic{equation}}

\section{Cavity optomechanical readout out of atomtronic SQUID}

We consider an atomtronic quantum interference device (AQUID) that consists of two Josephson junctions ($\mathrm{JJ}_1$ and $\mathrm{JJ}_2$). The JJs are moved towards each other to generate a bias current $I_{\mathrm{bias}}$, which results in the Josephson tunneling of atoms. To emulate the effect of an external magnetic field in the case of SQUID, we rotate the JJs around the ring with frequency $\Omega$. In this case, due to the single-valuedness of the wave function, the phase should satisfy~\cite{suzuki2006lecture,ryu2020quantum}
\begin{equation}
    \int \Delta \theta . dl = \theta_{2_x} - \theta_{1_x} + \theta_{1_y} - \theta_{2_y} = 2 \pi \frac{\Omega}{\Omega_0}.
\end{equation}
By defining phase difference across $\mathrm{JJ}_{1,2}$ as $\delta_{1,2}$ and the normalized rotation rate $\Omega' = \Omega/\Omega_0$, where $\Omega_0 = \hbar/mR^2$, we can write 
\begin{equation}
    \delta_1 = \delta_0 + \pi \Omega', \delta_2 = \delta_0 - \pi \Omega'.
\end{equation}
Using the above expressions, the total current in the AQUID is calculated by
\begin{equation}
    \begin{aligned}
        I = I_1 + I_2 &= I_c[\sin(\delta_1) + \sin(\delta_2)]\\
        & = I_c[\sin(\delta_0 + \pi \Omega') + \sin(\delta_0 - \pi \Omega')]\\
        & = 2I_c \cos(\pi \Omega') \sin(\delta_0).
    \end{aligned}
    \label{eq:SQUID_current}
\end{equation}
From Eq.~(\ref{eq:SQUID_current}), it is clear that the critical current of the AQUID $|2I_c \cos(\pi \Omega')|$ varies periodically with the barrier rotation rate $\Omega$ with period $\Omega_0$ and this modulation is a clear signature of interference of currents in the two arms of AQUID.

\begin{figure}[!htb]
\begin{center}
	\includegraphics[width= 0.3\linewidth]{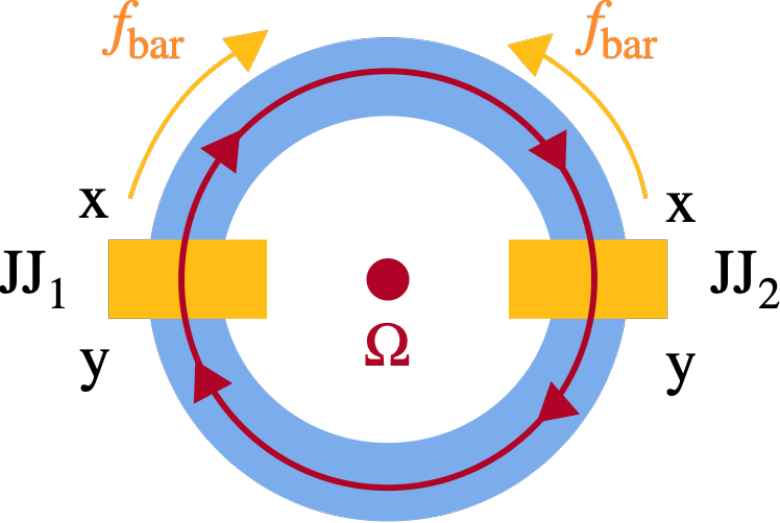} 
\end{center}
\caption{ Schematic diagram of an atomtronic quantum interference device (AQUID) that contains two Josephson junctions ($\mathrm{JJ}_1$ and $\mathrm{JJ}_2$) with rotation frequency $\Omega$. The JJs are moved towards each other with velocity $f_{\mathrm{bar}}$ to generate the chemical potential difference (bias voltage) and a bias current.  }
\label{fig:squid_schematic}
\end{figure}

\begin{figure}[!htb]
\begin{center}
	\includegraphics[width= 0.6\linewidth]{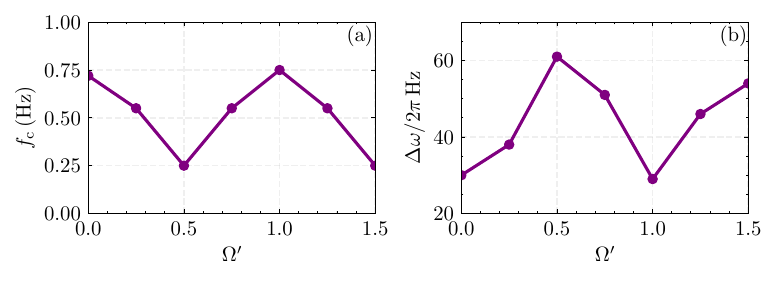} 
\end{center}
\caption{ (a) Variation of critical barrier velocity $f_{\mathrm{c}}$ with AQUID rotation rate $\Omega'$ and (b) variation of magnitude of peak splitting $\Delta \omega$ for $f_{\mathrm{bar}} = 0.75 \, \mathrm{Hz}$ and . }
\label{fig:int_Ic}
\end{figure}

\begin{figure}[!htb]
\begin{center}
	\includegraphics[width= 1\linewidth]{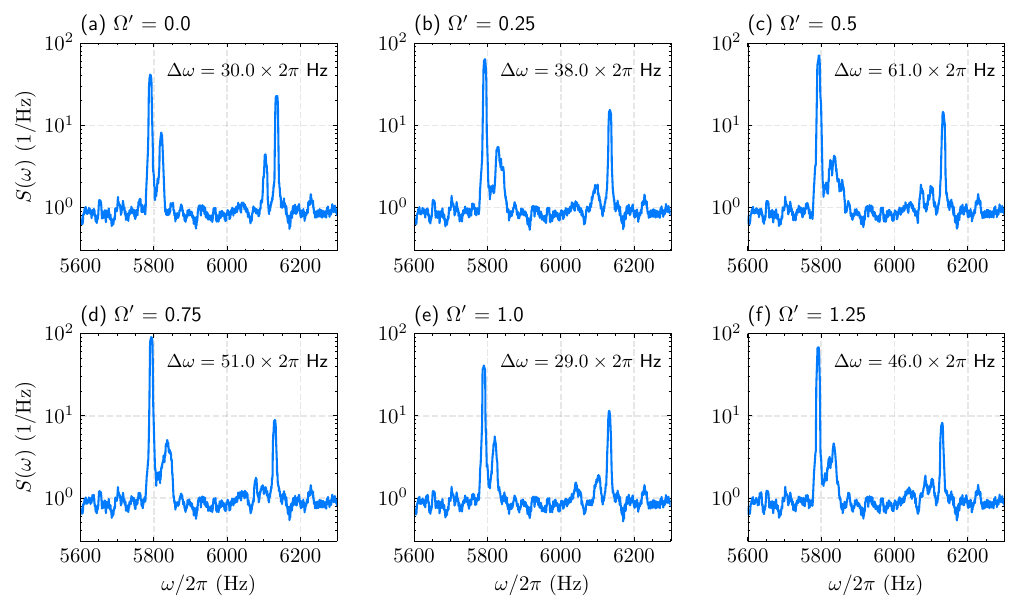} 
\end{center}
\caption{Power spectrum of the phase quadrature of the cavity output field for junction rotation frequency (a) $\Omega' = 0.0$, (b) $\Omega' = 0.25$, (c) $\Omega' = 0.5$, (d) $\Omega' = 0.75$, (e) $\Omega' = 1.0$, and (f) $\Omega' = 1.25$.  }
\label{fig:int_psd}
\end{figure}

To detect this interference, we plot the critical barrier velocity $f_{\mathrm{c}}$ for different rotation rates $\Omega'$ in Fig.~\ref{fig:int_Ic} (a). An easier way is to track the amount of split in the cavity output spectra for a certain barrier velocity, above $v_{\mathrm{c}}$ (See Fig.~\ref{fig:int_Ic} (b))  for different rotation rates $\Omega'$, which is similar to tracking the dc voltage across the SQUID as a function of external magnetic field~\cite{SQUIDbook}. We can see a clear modulation of the critical barrier velocity (i.e., critical current) and magnitude of peak splitting with the AQUID rotation rate with periodicity of one flux quantum of the ring $\Omega_0 = \hbar/ mR^2$.  The corresponding cavity output spectra are shown in Fig.~\ref{fig:int_psd}, demonstrating the quantum interference of currents.  

\section{Resistively and Capacitively Shunted Junction (RCSJ) Model}

Considering the non-zero capacitance of the Josephson junction and resistance due to quasi-particle excitation when the junction current is more than the critical current, the Resistively and Capacitively Shunted Junction (RCSJ) model is used to simulate the characteristics of a Josephson junction~\cite{Kiehn2022,bernhart2024observation, singh2024}. In this framework, using Kirchov’s circuit law, the total current flowing through a single junction can be written as

\begin{equation}
\begin{aligned}
        I &= I_J + I_R + I_C \\
        I &= I_c \sin(\Phi) - \frac{1}{R} \Delta \mu - C \frac{\Delta \mu}{dt}        
\end{aligned}
\label{eq:rcsj_single}
\end{equation}
Here, $I_J=I_c \sin(\Phi)$ is the current through the junction, $I_R=\frac{1}{R} \Delta \mu$ is the current through the resistor with resistance $R$, and $I_C = C \dot{\Delta \mu} $ is the current through the capacitor with capacitance $C$. $\Phi$ and $\Delta \mu$ are the phase difference and the chemical potential difference at the junction. Using the expression $\Delta \mu = -\hbar \dot{\Phi}$, Eq.~(\ref{eq:rcsj_single}) can be written as

\begin{equation}
    \begin{aligned}
        & \hbar C \ddot{\Phi} + \frac{\hbar}{R} \dot{\Phi} + I_c \sin(\Phi) = I \\
        \Rightarrow & \ddot{\Phi} +  \frac{1}{RC} \dot{\Phi} + \frac{I_c}{\hbar C} \sin(\Phi) = \frac{I}{\hbar C} \\
    \end{aligned}
\end{equation}

\begin{figure}[!htb]
\begin{center}
	\includegraphics[width= 0.5\linewidth]{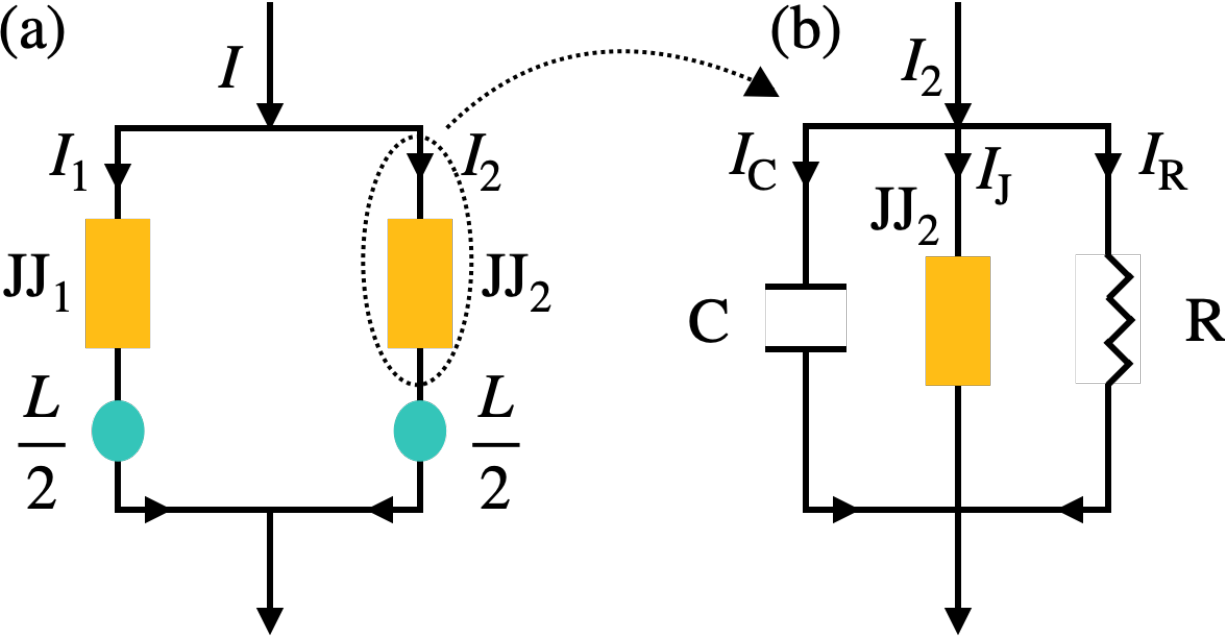} 
\end{center}
\caption{ (a) DC SQUID circuit with two Josephson junctions $\mathrm{JJ}_{1,2}$ and inductance $L$,  (b) RCSJ circuit for a single junction with resistance $R$ and capacitance $C$. }
\label{fig:squid_rcsj}
\end{figure}

Using the non-dimensionalization scheme $\tau = \omega_p t$, where $\omega_p = \sqrt{\frac{I_c}{\hbar C}}$, we arrive at
\begin{equation}
    \begin{aligned}
        & \frac{d^2 \Phi}{d \tau^2} \omega_p^2 + \frac{1}{RC} \frac{d \Phi}{d \tau} \omega_p + \omega_p^2 \sin(\Phi) =  \frac{I}{\hbar C} \\
        \Rightarrow & \boxed{ \frac{d^2 \Phi}{d \tau^2} + \frac{1}{Q} \frac{d \Phi}{d \tau}  +  \sin(\Phi) =  \frac{I}{I_c},}
    \end{aligned}
    \label{eq:rcsj_phi}
\end{equation}
where $Q = \omega_p RC$ is the quality factor, and the voltage across the junction is 
\begin{equation}
    V = \omega_p \left\langle \frac{d \Phi}{d \tau} \right\rangle. 
\end{equation}

For the AQUID device with two junctions, considering the external rotation and flux quantization condition of the toroidal geometry, the circuit can be described by
\begin{equation}
    \begin{aligned}
        &\frac{d^2 \Phi_1}{d \tau^2} + \frac{1}{Q} \frac{d \Phi_1}{d \tau}  +  \sin(\Phi_1) =  \frac{I_1}{I_c},\\
        &\frac{d^2 \Phi_2}{d \tau^2} + \frac{1}{Q} \frac{d \Phi_2}{d \tau}  +  \sin(\Phi_2) =  \frac{I_2}{I_c},\\
    \end{aligned}
\end{equation}

\begin{figure}[!htb]
\begin{center}
	\includegraphics[width= 1\linewidth]{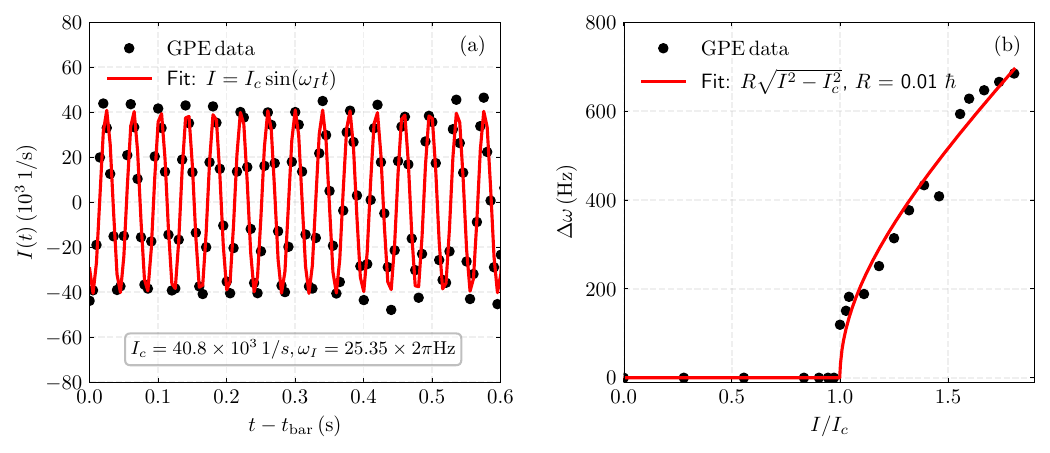} 
\end{center}
\caption{Fitting procedure to obtain $R$, $C$, and $I_c$. (a) The tunneling current $I = dN/ dt$ obtained from the GPE simulation is fitted by a sine function $I = I_c \sin (\omega_I t)$, which yields $I_c = 40.8 \times 10^3 \, \mathrm{1/s}$. (b) Magnitude of split in the peaks of the cavity output spectra $\Delta \omega$ is fitted by $ \hbar \Delta \omega = \Delta \mu = R \sqrt{I^2 - I_c ^2} $, which yields $R = 0.01 \hbar $. We obtain $C$ by calculating $n(t)/ \Delta \mu (t)$ \cite{bernhart2024observation} whose mean value gives $C = 4.33 \, \mathrm{s} / \hbar$.    }
\label{fig:fit_I_cR}
\end{figure}

\begin{figure}[!htb]
\begin{center}
	\includegraphics[width= 1\linewidth]{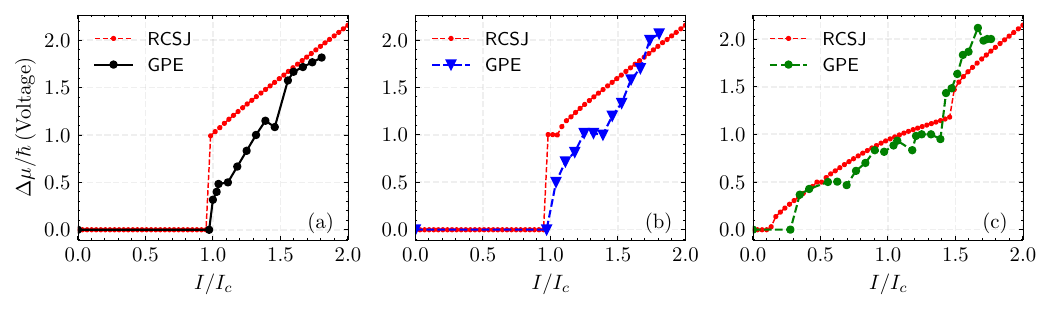} 
\end{center}
\caption{ Comparison between the results obtained from the GPE and RCSJ model for (a) the undriven case, (b) driven with $\Omega' = 0$ (that gives an integer Shapiro step), and (c) driven with $\Omega' = 0.5$ (that gives a half-integer Shapiro step). The set of parameters used here is the same as used in Fig. 4 of the main text.}
\label{fig:fit_I_cR}
\end{figure}
Defining $I_{1,2}/I_c = k_{1,2}$, $k_1 + k_2 = k$, and $k_2 - k_1 = 2k_s$, and $1/Q = \beta_j$, and using the flux quantization condition
$$2 \pi n - \Phi_1 - \Phi_2 - 2\pi (\Omega' + \frac{\beta}{2} k_s)$$, where $\beta = 2 \pi I_c/ N \Omega_0$, we obtain the equations
\begin{equation}
\boxed{
\begin{aligned}
    &\frac{d^2 \Phi_1}{d \tau^2} + \beta_j \frac{d \Phi_1}{d \tau}  +  \sin(\Phi_1) =  \frac{k}{2} - k_s,\\
    &\frac{d^2 \Phi_2}{d \tau^2} + \beta_j \frac{d \Phi_2}{d \tau}  +  \sin(\Phi_2) =  \frac{k}{2} + k_s,\\
    &k_s = \frac{2}{\beta} \left[ \frac{\Phi_1 - \Phi_2}{2\pi} - \Omega' - R \left(\frac{\Phi_1 - \Phi_2}{2\pi} - \Omega' \right) \right],
\end{aligned}
}
\label{eq:rcsj_driving}
\end{equation}
where $R \Big(\frac{\Phi_1 - \Phi_2}{2\pi} - \Omega' \Big)$ is the nearest integer function and its subtraction yields the deviation from the quantized integer winding number, that drives the screening current. 

To account for the periodic driving of the barrier ($\phi_{b_{1,2}} = \pm (2 \pi f_{\mathrm{bar}} ) t + \phi_i \, \sin(2 \pi f_{\mathrm{m}})$) that generates the Shapiro steps, we have added the modulation current $\frac{I_m}{I_c} \sin (\frac{2 \pi f_m \tau}{\omega_p})$ in Eq.~\ref{eq:rcsj_driving}, where $f_m$ is the modulation frequency and $\frac{I_m}{I_c} = \frac{2 \pi f_m \times \phi_I}{V_c}$. The voltage across the AQUID is calculated via
\begin{equation}
    V = \frac{\omega_p}{2}  \left\langle \frac{d \Phi_1}{d \tau}+ \frac{d \Phi_2}{d \tau}  \right\rangle. 
\end{equation}

\section{Characteristics of the Shapiro steps}

\subsection{Effect of the barrier rotation}

\begin{figure}[!htb]
\begin{center}
	\includegraphics[width= 1\linewidth]{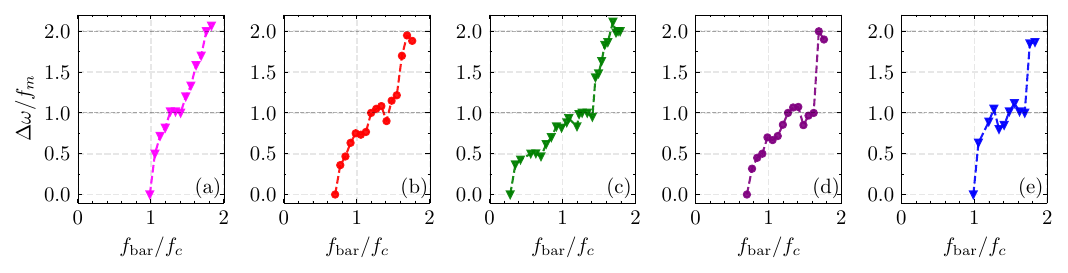} 
\end{center}
\caption{Shapiro steps for barrier modulation amplitude $\phi_I = 0.004$ and barrier rotation frequency (a) $\Omega' = 0.0$, (b) $\Omega' = 0.25$, (c) $\Omega' = 0.5$, (d) $\Omega' = 0.75$, and (e) $\Omega' = 1.0$. The other set of parameters used here is the same as used in Fig. 2 of the main text.    }
\label{fig:shapiro_rot}
\end{figure}

In this section, we will present the variation of the amount of peak splitting ($\Delta \omega$) in the cavity output spectra as a function of the barrier velocity $f_{\mathrm{bar}}$ for different barrier rotation frequencies $\Omega'$ in Fig.~\ref{fig:shapiro_rot}. For $\Omega' = 0$, the case of zero rotation, we can see the integer Shapiro steps, which were seen earlier in the rectangular BEC cloud~\cite{bernhart2024observation}. When we start rotating the junctions, we can see the emergence of additional steps whose height is around $0.75$ (Fig.~\ref{fig:shapiro_rot} (b)), and the first step height becomes exactly $0.5$ for $\Omega' = 0.5$, which is the half-integer Shapiro steps (See Fig.~\ref{fig:shapiro_rot} (c)). By further increasing the rotation frequency, the half-integer step vanishes and the integer step emerges for $\Omega' = 1$ (Fig.~\ref{fig:shapiro_rot} (e)). So we get the integer Shapiro steps for integer flux quantum and the half-integer steps for half-integer flux quantum. 

\begin{figure}[!htb]
\begin{center}
	\includegraphics[width= 1\linewidth]{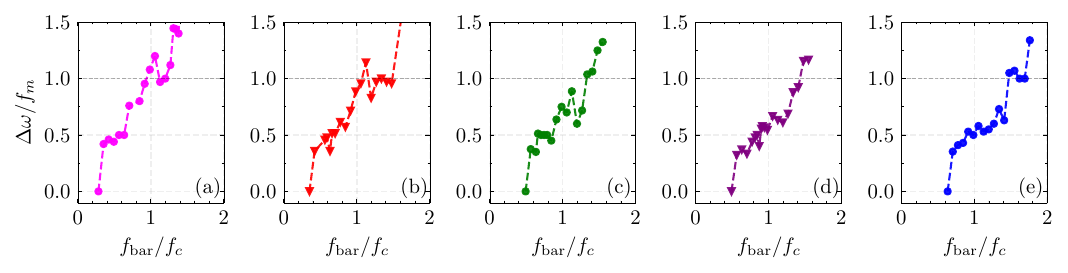} 
\end{center}
\caption{Shapiro steps for barrier modulation amplitude $\phi_I = 0.004$ and barrier modulation frequency (a) $f_m = 50$ Hz, (b) $f_m = 70$ Hz, (c) $f_m = 80$ Hz, (d) $f_m = 90$ Hz, and (e) $f_m = 100$ Hz. The other set of parameters used here is the same as used in Fig. 2 of the main text.  }
\label{fig:shapiro_fm}
\end{figure}

\subsection{Effect of the barrier modulation}

In this section, we will present the variation of the amount of peak splitting ($\Delta \omega$) in the cavity output spectra as a function of the barrier velocity $f_{\mathrm{bar}}$ for different modulation frequencies $f_{m}$ and $\Omega' = 0.5$ in Fig.~\ref{fig:shapiro_fm}. As discussed in Fig.~\ref{fig:shapiro_rot}, we get the half-integer Shapiro steps for half flux quantum $\Omega'=0.5$ and we demonstrate it for a range of modulation frequency $f_m$. For all the cases, we get the half-integer Shapiro steps.

\subsection{Cavity output spectra corresponding to Integer and half-integer Shapiro steps}

\begin{figure}[!htb]
\begin{center}
	\includegraphics[width= 1\linewidth]{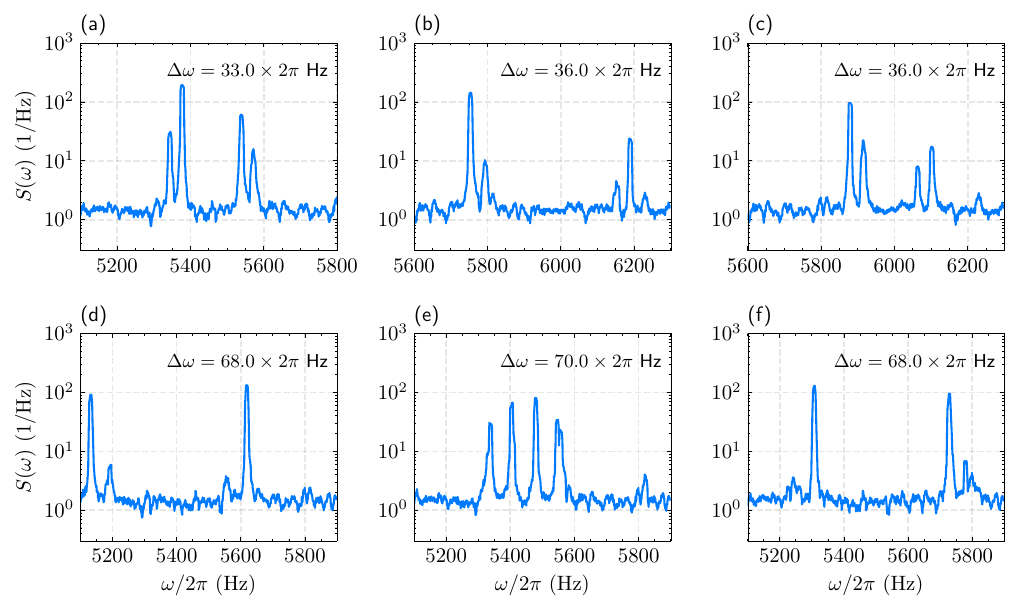} 
\end{center}
\caption{Power spectrum of the phase quadrature of the cavity output field for barrier rotation rate $\Omega' = 0.5$, barrier modulation frequency $f_{\mathrm{m}} = 70\, \mathrm{Hz}$, modulation amplitude $\phi_I = 0.004$, and barrier velocity (a) $f_{\mathrm{bar}} = 0.42 \, \mathrm{Hz}$, (b) $f_{\mathrm{bar}} = 0.47 \, \mathrm{Hz}$, (c) $f_{\mathrm{bar}} = 0.5 \, \mathrm{Hz}$, (d) $f_{\mathrm{bar}} = 0.9 \, \mathrm{Hz}$, (e) $f_{\mathrm{bar}} = 0.95 \, \mathrm{Hz}$, and (f) $f_{\mathrm{bar}} = 1.0 \, \mathrm{Hz}$. The upper row corresponds to the half-integer Shapiro steps, and the lower row corresponds to the integer Shapiro steps.  }
\label{fig:psd_fm70}
\end{figure}

In this section, we present the power spectrum of the phase quadrature of the cavity output field for different barrier velocities $f_{\mathrm{bar}} $, keeping the barrier rotation rates and barrier modulation frequency constant in Fig.~\ref{fig:psd_fm70}. For (a) $f_{\mathrm{bar}} = 0.42 \, \mathrm{Hz}$, (b) $f_{\mathrm{bar}} = 0.47 \, \mathrm{Hz}$, (c) $f_{\mathrm{bar}} = 0.5 \, \mathrm{Hz}$, we can see the magnitude of splitting of the peaks is the same, and they are nearly half of the barrier modulation frequency $f_{\mathrm{m}} = 70\, \mathrm{Hz}$. These values contributes to the half-integer Shapiro steps in Fig.~\ref{fig:shapiro_fm} (b). For (d) $f_{\mathrm{bar}} = 0.9 \, \mathrm{Hz}$, (e) $f_{\mathrm{bar}} = 0.95 \, \mathrm{Hz}$, and (f) $f_{\mathrm{bar}} = 1.0 \, \mathrm{Hz}$, the magnitude of splitting of the peaks is nearly same as the barrier modulation frequency $f_{\mathrm{m}} = 70\, \mathrm{Hz}$ and contributes to the integer Shapiro steps in Fig.~\ref{fig:shapiro_fm} (b).

\end{document}